# Optical Valley Hall Effect of 2D Excitons in Hyperbolic Metamaterial

**Sriram Guddala[1], Mandeep Khatoniar[1,2], Nicholas Yama[1], W. Liu[3,#], G. S. Agarwal[3], and Vinod M. Menon[1,2, *]**

[1]Department of Physics, City College of New York, City University of New York, New York 10031, United States

[2]Department of Physics, The Graduate Center, City University of New York, New York 10016, United States

[3]Institute for Quantum Science and Engineering, and Department of Biological and Agricultural Engineering ; Texas A&M University, College Station, Texas 77843, USA

**The robust spin and momentum valley locking of electrons in two-dimensional semiconductors makes the valley degree of freedom of great utility for functional optoelectronic devices. Owing to the difference in optical selection rules for the different valleys, these valley electrons can be addressed optically. The electrons and excitons in these materials exhibit valley Hall effect, where the carriers from specific valleys are directed to different directions under electrical or thermal bias. Here we report optical valley Hall effect where the light emission from the valley polarized excitons in monolayer $WS_2$ propagates in different directions owing to the preferential coupling of excitonic emission to the high momentum states of the hyperbolic metamaterial. The experimentally observed effects are corroborated with theoretical modeling of excitonic emission in the near field of hyperbolic media. The demonstration of optical valley Hall effect using a bulk artificial photonic media without the need for nanostructuring opens the possibility of realizing valley based excitonic circuits operating at room temperature.**

---------------------------------------------------------------------------------------------------------------

**\* Corresponding Author:** <vmenon@ccny.cuny.edu>

**# Current Affiliation:** Shaanxi Province Key Laboratory for Quantum Information and Quantum Optoelectronic Devices, and Department of Applied Physics, Xi'an Jiaotong University, Xi'an 710049, China

## I. Introduction

In the monolayer limit, transition metal dichalcogenides (TMDs) such as $WS_2$, $WSe_2$, $MoSe_2$, and $MoS_2$ are direct-bandgap semiconductor materials with broken inversion symmetry[1–3]. These materials exhibit very strong interaction with light absorbing greater than 10% light at the monolayer limit close to the excitonic resonances. The excitons in these materials have unusually large binding energies (0.1- 0.5 eV), can be developed into functional optoelectronic devices such as light emitting diodes and detectors[4–6], and even demonstrate strong coupling to cavity photons at room temperature[7,8]. The direct bandgap in these materials occurs at the *K* point in the momentum space in the monolayer limit. An intriguing property that arises from the broken inversion symmetry and the strong spin-orbit interaction arising from the heavy transition atoms is valley polarization - quantum mechanically distinct valleys *K* and *K'* which can be selectively addressed by right handed and left handed circularly polarized light as shown schematically in Fig. 1a. Valley polarization has been studied extensively in these systems and has been touted as a path to encode and manipulate information[9–11] resulting in the emerging field of valleytronics[12,13]. Limiting factors for practical applications of valley degree of freedom arises from the intervalley scattering and disorder induced depolarization. Enhancement of valley polarization via strong coupling to cavity photons as well as their routing using nanophotonic structures has recently garnered much attention and promise for realization of functional valleytronic devices[14–18]. In all these demonstrations, the enhancement of valley degree of freedom was achieved via optical cavities or very specific resonant plasmonic and dielectric nanostructures. More recently the optical valley Hall effect was demonstrated in the context of strongly coupled exciton polaritons[19]. The optical valley Hall effect can be thought of as an optical analog of the conventional Hall effect which in the context of valley degree of freedom is termed as valley Hall effect, which results in directional propagation of carriers under electric field[20,21]. Excitonic valley Hall effect using thermal gradient produced by an excitation laser was also recently demonstrated [22].

Here we report the use of hyperbolic media to realize optical valley Hall effect where the 2D TMD does not have to be placed inside cavities or on any nanostructured medium. Instead the near field coupling of the valley polarized excitonic emission to high-k states of the hyperbolic media facilities the optical valley Hall effect resulting in directional propagation of the emission. This optical valley Hall effect in hyperbolic media arises due to the nature of light propagation inside such medium and has previously been used to demonstrate optical spin Hall effect in the microwave[23]. While we do find weak optical valley Hall effect even with a simple metallic film, the observed valley polarization contrast is much stronger and has higher contrast in the case of the multilayered hyperbolic metamaterial (HMM). We use momentum and helicity resolved spectroscopic techniques to demonstrate the optical valley Hall effect where we observe valley selective routing of excitonic emission in 2D $WS_2$ integrated with HMM. The observed phenomena are modeled using both numerical and analytical techniques and agrees with the experimental observation. The HMM essentially acts as an extremely broadband valley-beam splitter for circularly polarized valley exciton emission and a route to develop valley filters or valley-based signal processing circuits.

## II. Results

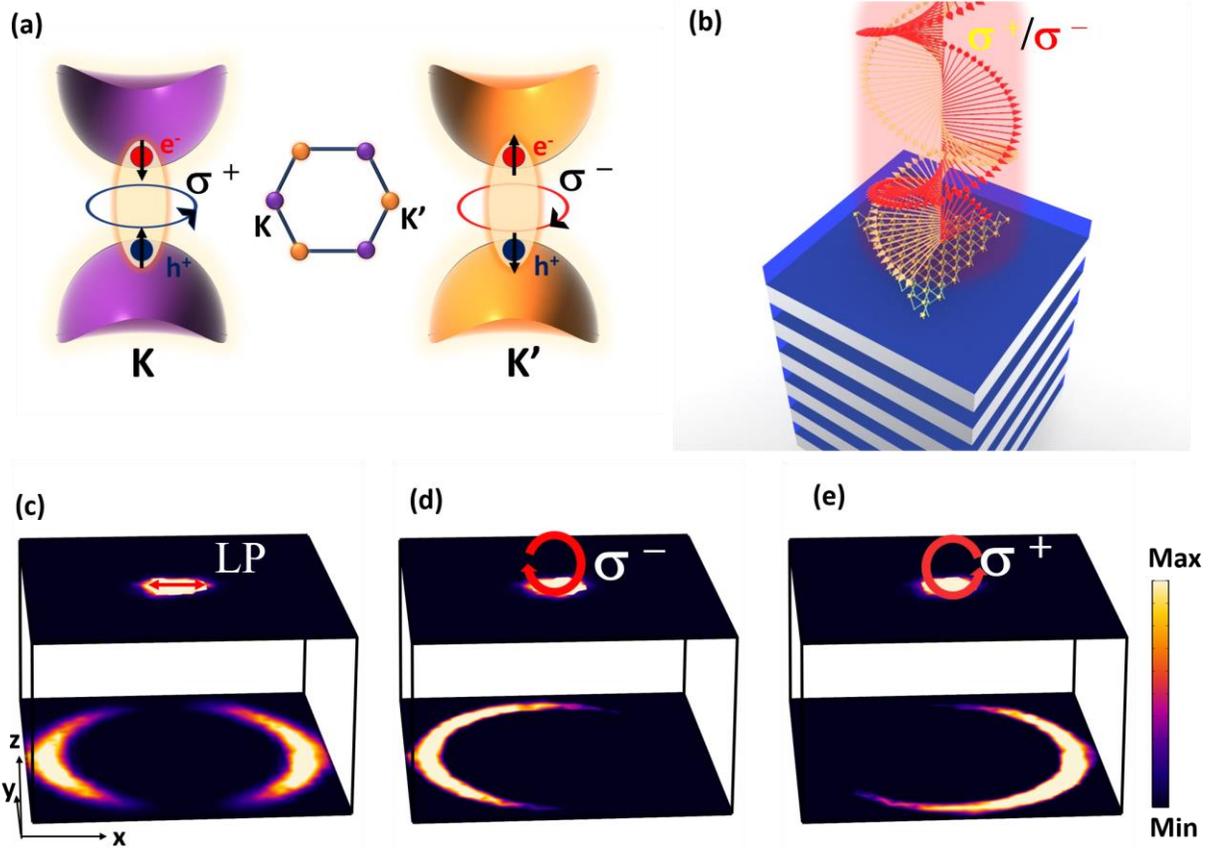

**Figure 1: Optical valley Hall effect demonstration in HMM.** a) schematic of Valley polarization in $WS_2$ monolayer, where *K* and *K'* valleys can be populated independently with $\sigma^+$ and $\sigma^-$ polarizations respectively. (b) Schematic of the seven periods HMM composed of $Ag/Al_2O_3$ shows a monolayer of $WS_2$ transferred to its surface layer. Electric field intensity distribution of the high *k*-modes for the dipole radiating in the near field of HMM surface for (c) linear polarization and (d) and (e) are right- and left-handed circular polarizations respectively.

The artificial optical media was designed to have type II hyperbolic (single sheet hyperbola) iso-frequency dispersion in the visible region of the electromagnetic spectrum[24]. A schematic illustration of the HMM is shown in Fig .1b. The structure consisted of alternating layers of Ag and $Al_2O_3$ layers and the thickness of each layer was optimized to be 10 nm. The effective dielectric constants of the HMM simulated using effective medium theory is shown in the Supplementary Materials (Fig. S1(a)). Such HMMs have been previously used to enhance spontaneous emission from emitters owing to the enhanced local photonic density of states.[25–29]

Finite element method simulations were performed to investigate near field coupling of linear and circular polarization emissions to plasmon modes of the HMM. The HMM is taken to have effective dielectric permittivity values of $\epsilon_x = \epsilon_y$= -6.246+$i$1.081 and $\epsilon_z$ = 7.76+$i$0.2624 at wavelength of 620 nm as obtained from effective medium theory calculations. As shown in Figure 1(c), a linear dipole with dipole moment P$_{2D}$= $p$[1, 0] with unit polarization $p$ along the x-axis shows characteristic high-$k$ mode propagation in the HMM medium with symmetric electric field distribution in the $xy$-plane with symmetry about the y-axis. For right- and left-handed circular polarizations, two linear dipoles were assumed about the origin one in $xy$ plane and another in $xz$ plane with phase difference of ±90º. The sign of the phase difference defines the right and left handedness of the generated circular polarization. A right-handed circularly polarized dipole (σ $^-$) with, P$_{2D}$= $p(x-iz)$ /$\sqrt{2}$ is initially assumed in the $xz$ plane and swept for all azimuthal rotations about the z-axis and the resultant electric field distribution from all azimuthal rotations is plotted in Fig. 1(d). The high-$k$ mode electric field distribution is found to be asymmetric and strictly unidirectional in the -$x$ direction as shown in Figure 1(d). Similarly, this high-$k$ mode dispersion is towards +$x$ direction for left-handed circular polarization (σ $^+$) with, P$_{2D}$= $p(x+iz)$ /$\sqrt{2}$ as shown in Fig. 1(e). The fundamental principle behind the unidirectional excitation of surface plasmon mode and high-$k$ modes is governed by the selective vectorial excitation of the electric field associated with these modes. The handedness dependent unidirectional coupling of chiral fields to the HMM can be thought of as the photonic analogue of spin-Hall like effect for photons which was previously demonstrated in pioneering experiments for surface plasmons in the metals[23,30–32] and RF hyperbolic metamaterials and more recently shown theoretically for chiral emitters[33]. TMDs owing to their strong valley polarization property emits circular polarized photoluminescence (PL) when excited with the same handedness owing to the valley specific excitation ($K$ or $K$'). Thus, an HMM-TMD integrated system naturally lends itself to demonstrate optical analogue of valley-Hall like effect for valley polarized emission of the excitons in TMDs and provides a path towards integrated valleytronics.

To study the problem of chiral emission through HMM, we assume a two-dimensional electric dipole source located on the upper surface of a multilayer structure. The electric field density in the lower half-space can be calculated by the simplified expression $E_x \propto (\alpha + \beta k_x/|k_z|)$. Here, $k_x$ and $k_z$ are the x and z components of the wavevectors and $\alpha$, $\beta=0, \pm 1$ are constants that define the electric dipole polarization. Note that the expression for electric field includes the contribution of the propagating [$\sqrt{k_x^2 + k_y^2} < k_0$] and evanescent [$\sqrt{k_x^2 + k_y^2} > k_0$] modes as shown by us previously[33]. The latter term is responsible for near field interference effects which is central to the chiral selectivity observed in HMMs as discussed below.

If $\alpha > 0$ and $\beta > 0$, the spectral amplitude with $k_x < -k_0$ add up destructively, whereas for $k_x > k_0$ constructive interference occurs. Therefore, the near-field interference effect lies at the heart of the selective directional excitation of guided modes illuminated by an elliptically polarized dipole. When an elliptically polarized dipole illuminates a multilayered structure, the inversion symmetry is broken, and the high-$k$ modes can be excited. The high-$k$ evanescent modes carry transverse spin angular momentum[31,32,34]. Therefore, the best coupling is achieved when the evanescent mode with transverse spin matches the same spin (helicity) of the incident/emitted field. By tuning the polarization of the dipole, the excitation and guidance of electromagnetic modes in a nanostructure can be controlled.

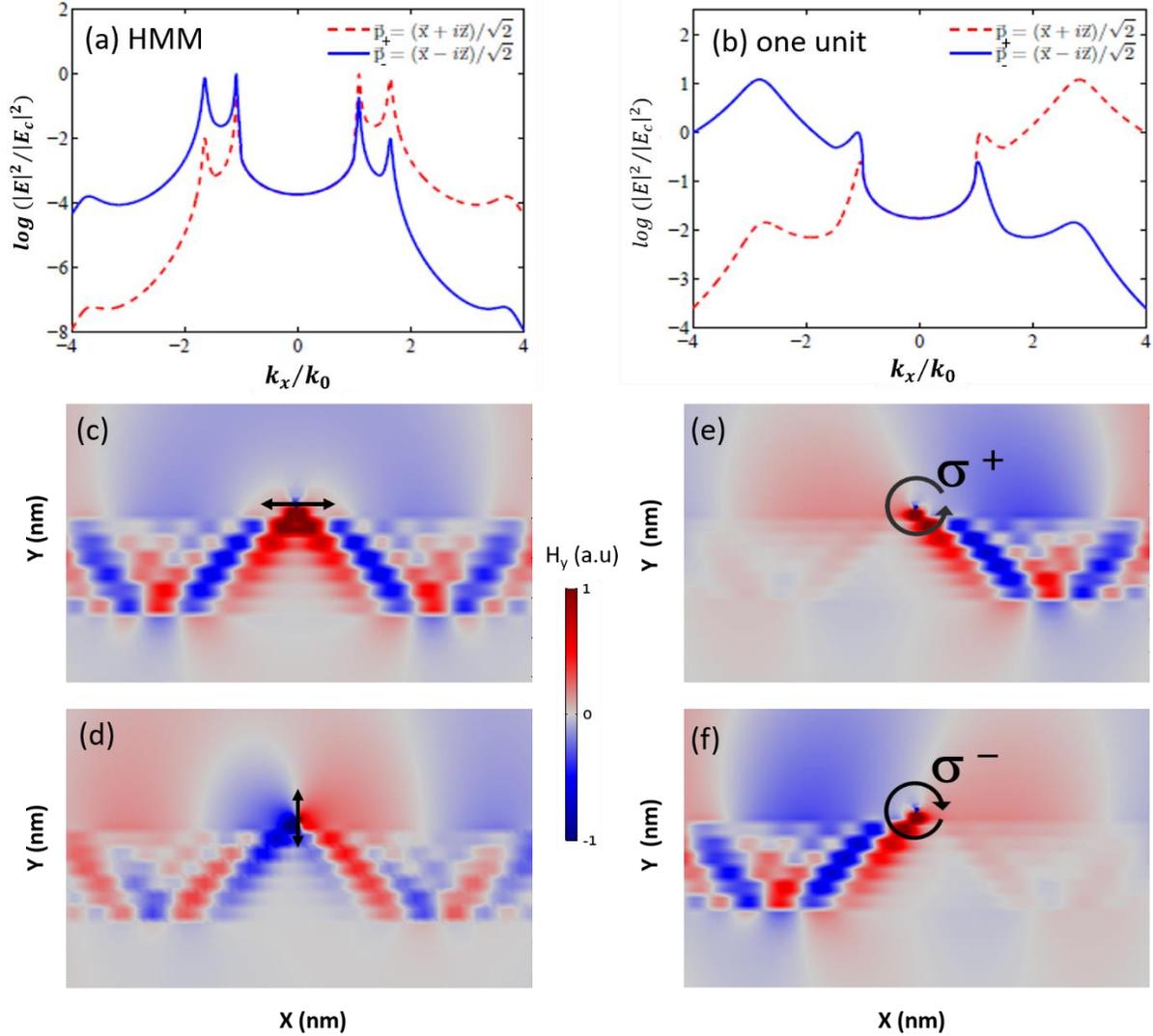

**Figure 2: Vectorial field of high-$k$ modes in HMM.** Mode spectrum calculated from TMM simulations for circular dipoles with opposite helicities in the *xz* plane of a) 7P HMM sample and b) one unit HMM. The magnetic field $H_y$ component intensity distribution in the HMM for linear dipole with polarization orientation along: (c) horizontal P = $\hat{x}$ and (d) vertical P = $\hat{z}$. (e) Left-handed circularly polarized dipole P= $(\hat{x}+i\hat{z})/\sqrt{2}$ and (f) right-handed circularly polarized dipole P= $(\hat{x}-i\hat{z})/\sqrt{2}$. These simulations were carried out at the emission wavelength of the exciton (620 nm).

Using this theoretical model and finite element method simulations we show how to control the excitation direction of electric field by tuning the polarization of the dipole emitter. The unit vectors for the horizontally and vertically polarized dipoles are $p_x = x$ and $p_z = z$, respectively. For circularly polarized dipoles, we have unit vectors $p_\pm = (\alpha x + i\beta z)/\sqrt{2}$ with $\alpha = 1$ and $\beta = \pm 1$. HMMs support two kinds of modes namely the surface plasmon polaritons (SPPs) at the interfaces with the claddings and the high-$k$ modes confined to the bulk of the layered material[28,29,35]. The SPP modes in HMM behave fundamentally different than in a single thin metallic film due to fact that these modes are resultant of mutual repulsion of SPP modes at each metal-dielectric interface. The large values of model index for the SPPs (called long and short range plasmon polaritons) in HMM medium results in higher field confinement in comparison SPPs in single metallic film[35]. As a result, the SPP modes in layered medium show much reduced line widths than the surface modes of a thin film as shown in Fig.2a. For a linear dipole located at the top surface of the HMM is well-known to show SPPs and high-$k$ wavevector modes[28,29,35]. The TMM simulations for our seven period HMM show a symmetric dispersion of mode spectrum (supporting information Fig. S2). Whereas the circular polarizations show asymmetric dispersion as shown in Fig. 2a for σ$^+$ (red) and σ$^-$ (blue). For a thin metallic film with a dielectric corresponding to one-unit cell of the HMM, a broad SPP mode is observed (Fig. 2b). The asymmetric dispersion for SPP mode (at $k_x = \pm 1.057$) in HMM is found to have almost similar contrast as one unit HMM with unidirectional optical routing for specific chirality. Whereas the contrast supported by a low quality-factor SPP mode in thin metallic layer is experimentally found to be indistinguishable. The electric field patterns for linear and circular polarizations are shown in Fig.2. Distinct asymmetric field pattern is observed for a circular dipole (Fig. 2e and 2f) in comparison to the horizontal (Fig.2c) and vertical (Fig.2d) dipoles. Since a circular dipole is a linear combination of horizontal and vertical dipoles, the electric field pattern for the circular dipole can be interpreted as linear combination of the linear dipoles. It can be noticed from the phase relations of the fields generated by the horizontal and vertical dipoles that these fields can be summed up, and the qualitative result can be immediately inferred to observe the unidirectional propagation of the modes of the HMM for circular polarization dipole excitation. Finite element simulations were also carried out for one-unit cell of the HMM consisting of a thin silver film with a dielectric layer. Fig. S3 shows σ$^+$ and σ$^-$ circular dipole emissions for the one-unit cell of the HMM. It is clear from these simulations that for the same circular dipole emission, HMM shows more well-defined unidirectional coupling in comparison to the one period HMM.

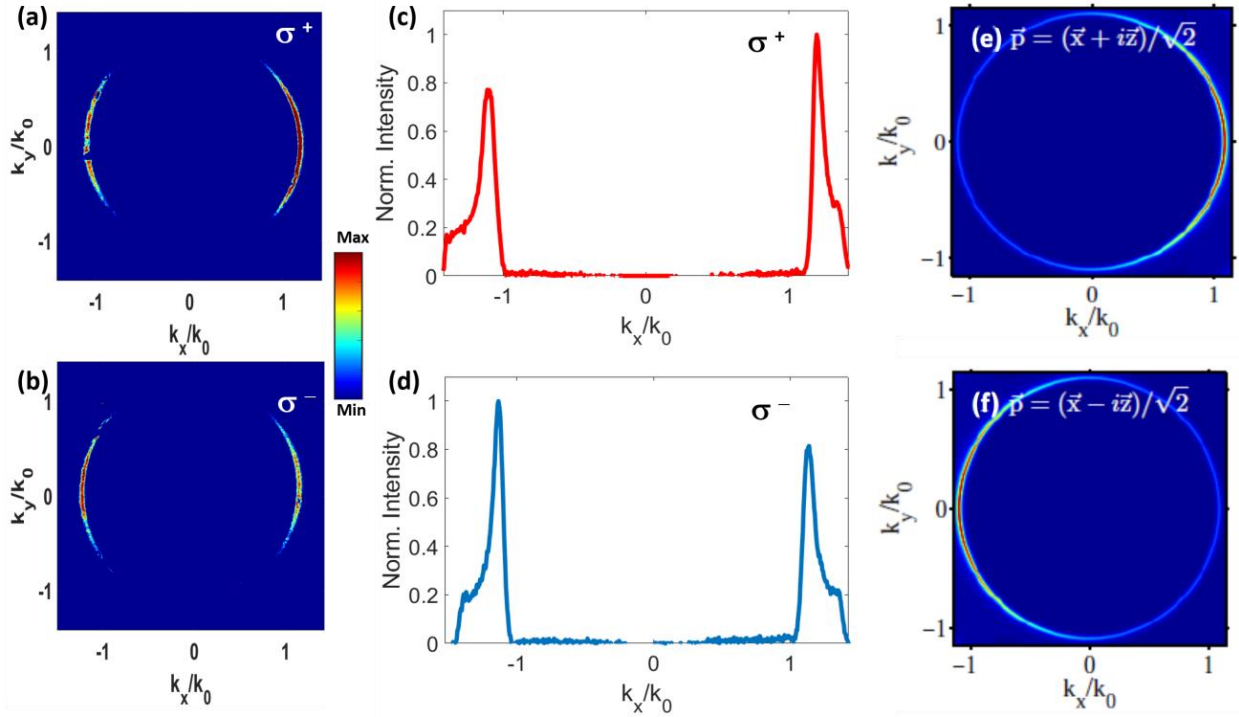

**Figure 3: Fourier space imaging of optical Valley- Hall effect in HMM.** Fourier space images of valley emissions under helicity resolved condition show the asymmetric intensity distribution for both a) $\sigma^+$ and (b) $\sigma^-$. (c) and (d) are the line cut profiles along the $k_x$ showing the asymmetric intensity for both $\sigma^+$ (a) and $\sigma^-$ (b) respectively. (e) and (f) show simulated Fourier space images for both $\sigma^+$ and $\sigma^-$ excitations respectively.

To demonstrate optical valley Hall effect of $WS_2$ monolayer, two energy degenerate $K$ and $K'$ valley emissions need to be directed in two opposite directions. Since the $K$ and $K'$ Valleys of the $WS_2$ emit light with specific handedness, the near field interaction of $K$ and $K'$ valley emissions with high-$k$ and SPP modes of HMM can be selectively routed to $+x$ and $-x$ directions, respectively. Thus, the optical valley Hall effect in monolayers can be mediated through the robust optical spin Hall effect supported by the modes of the HMM. Since the optical spin Hall effect in HMM is broadband, the optical valley Hall effect in the HMM also can be broadband over its broad hyperbolic regime unlike the earlier studies using resonant nanostructured media [15–18].

For the experimental verification of optical valley Hall effect, we fabricated a 140nm thick seven period HMM structure on 120μm thick glass slide through sequential depositions of Ag and $Al_2O_3$ thin films each of 10 nm thickness with 1 nm thick Ge wetting layer for each Ag layer using electron beam evaporation. A TMD monolayer of $WS_2$ was exfoliated to PDMS stamp and transferred to the HMM surface following our earlier reported process[18,36]. Helicity resolved k-space measurements were performed in transmission configuration using a home-built microscope set up (Fig. S4). The polarization of the incident laser was set to circular polarization using a combination of linear polarizer and quarter waveplate. The collection path polarization was resolved into σ⁺ and σ⁻ helicities using another set of quarter waveplate and linear polarizer. The monolayer was excited at exciton resonance wavelength of 620 nm from the top of the HMM surface with σ⁺ and σ⁻ polarizations using 10X microscopic objective with 0.25 NA. The emission from the bottom surface of the HMM is collected by using an oil immersion objective of 100X magnification with 1.4 numerical aperture. The back focal plane of the 100X objective is imaged by using Fourier space lens onto the CCD camera as shown in the experimental set up (Supporting information). The exciton emission was resolved to σ⁺ and σ⁻ emissions prior to Fourier space lens and finally imaged to the CCD camera. From Fig.3, it can be noticed that under σ⁻ excitation, the σ⁻ emission corresponding to K' valley excitons is found to be in the - $k_x$ direction. Whereas, σ⁺ emission from K valley for σ⁺ excitation is in +$k_x$ direction. A complete set of helicities resolved measurements are shown in the supporting information Fig. S5. These asymmetric intensity distributions in the *k*-space images can also be noticed from the line cuts made along the *kx* direction in the *k*-space images for σ⁺ and σ⁻ emissions in Fig. 3(c) and 3(d) respectively. The experimentally observed directional coupling of K and K' valley emissions to SPP are similar to the mode spectrums shown in Fig.2a. The simulated Fourier space images for σ⁺ and σ⁻ excitations in Fig. 3(e) and 3(f) were obtained by performing the Fourier transform over the field distributions at the bottom plane of the HMM medium from TMM simulations for chiral dipoles. The experimental observations agree with simulated results and indicates the routing of emission from the WS$_2$ excitons using HMMs. Additionally, the PL emission is observed only at SPP mode owing to the experimentally permitted modes of the HMM by the available oil immersion objective for the visible region. However, this asymmetric dispersion is true for both SPPs and high-*k* modes as shown in the simulations in Fig. 2. The phenomenon of unidirectional routing for σ⁺ and σ⁻ is same for both SPP and high-*k* modes, however, only the experimentally accessible ($k_{max}$ ~1.4) SPP modes were measured in this study.

A plasmonic thin film is also known for directional coupling of SPP wave for the incident chiral electromagnetic wave[34,37]. To compare the HMM to a planar silver film, we investigated the effect of directional coupling of WS$_2$ emission on one-unit cell of the HMM (10 nm silver film covered with 10nm Al$_2$O$_3$ film) and subsequently a monolayer of WS$_2$ transferred to its surface. Figure 4 shows helicity resolved k-space images for thin silver film sample for both σ⁻ and σ⁺ helicities. Thin silver film helicity resolved valley polarization measurements show (Fig. 4) much lower selectivity of valley polarized emissions compared to the 7 period HMM (Fig. 3). The helicity resolved valley emission measurements for one-unit cell HMM sample are shown in Fig. S6. The line cuts for both σ⁺ and σ⁻ excitations are shown in Fig. 4c and 4d respectively. The central spot at $k_x/k_0 = k_y/k_0 = 0$ is due to the incident laser intensity transmission through the thinner one-unit cell HMM sample. This is hardly observed in the case of seven period sample owing to the low direct transmission at *k*=0 and removed to show similarity to the simulation results. The valley specific emission directionality is less visible in the one-unit cell HMM sample. It is evident that seven period HMM shows (Fig. 3) promising valley contrast in comparison to thin film sample (Fig. 4a). The low contrast valley directionality is due to the fact that one-unit cell HMM supports low contrast mode confinement in the one-unit cell HMM simulations. It can also be understood by analyzing the simulated mode spectrum by using the electric field expression for chiral dipole located at surface of thin film sample. The mode spectrum of one-unit cell shows broad SPP modes as shown in Fig. 2b in comparison to the seven period HMM in Fig.2a. The broad line width in one-unit cell thick HMM is due to the low model index value of SPP in comparison to the seven period HMM medium SPPs model index value[35]. These factors could be the reason to support fast decaying SPP mode and hence lesser contrasting valley resolved directionality.

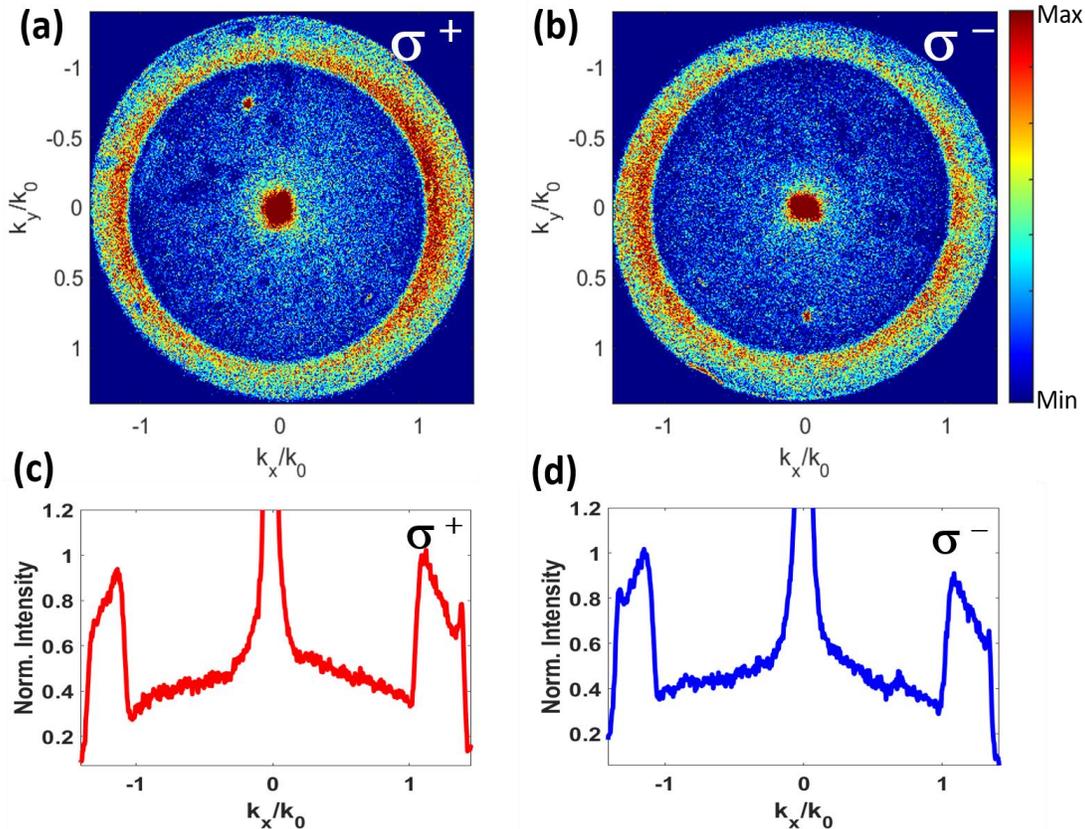

Figure 4: (a) and (b) show valley resolved $k$-space maps for one-unit cell HMM sample for both $\sigma^+$ and $\sigma^-$ helicities respectively. (c) and (d) show line cuts along the $k_x$ directions corresponding to $\sigma^+$ and $\sigma^-$ helicities respectively.

These experimental observations demonstrate that the $K$ and $K'$ valley excitonic emissions unidirectionally couple to the SPP mode of the HMM analogous to photonic spin-Hall effect.[23,34] The contrast between the intensity along $-k_x$ and $+k_x$ for both $\sigma^+$ ( and $\sigma^-$ ) is found to be more than 20%. This contrast for selective valley directional guiding is found to be significantly higher in comparison to the one layer of silver film.

In summary, we demonstrate efficient routing of valley exciton emission using hyperbolic media by exploiting the high-$k$ states and the higher Q factor SPP modes in them. This effect relies on the photonic spin Hall effect in HMMs. The experimental demonstration is found to be in very good agreement with theoretical calculations. Through Fourier space imaging we are able to establish the valley dependent routing of emission from monolayer WS$_2$ placed in the near field of the HMM. While nanostructured photonic systems can be used to realize similar effects, the use of a bulk substrate to realize such valley polarized routing presents a simple and broadband approach for practical realization of valley dependent optoelectronics.

## Methods

**Simulations**

Finite element method simulations were performed by using COMSOL Multiphysics RF module. The HMM is designed with effective medium parameters of $\varepsilon_x = \varepsilon_y = -6.246 + i1.081$ and $\varepsilon_x = 7.76 + i0.2624$ for Ag/Al$_2$O$_3$ multilayers volume fraction of 50%. The multilayered HMM is also designed with Ag/Al$_2$O$_3$ multilayers in the XY plane with periods extending in the z-direction. Simulations model the quantum emitter as a circular electric dipole oriented in the XZ plane perpendicular to the metal-dielectric layers of the HMM with an emission spectrum at 620 nm. The dielectric constant for the Ag is obtained from the previous report of Palik et al,[38]. The dielectric constant for the Al$_2$O$_3$ layer is considered as $\varepsilon_{Al_2O_3} = 1.67$. Transfer matrix method simulations were implemented to obtain HMM and one-unit HMM samples mode dispersions for both linear and circular dipole excitations.

**Sample fabrication.** Seven periods HMM substrate and one period control sample with Ag and Al$_2$O$_3$ multilayers were grown by a Kurt Lesker PVD 75 electron beam evaporation system on top of precleaned 120 μm thick glass substrates with surface roughness<1 nm. The pressure inside the vacuum chamber was kept to ~2× 10$^{-6}$ torr throughout the process. 1- 2 nm Ge seed layer was grown prior to each Ag layer to ensure good uniform distribution of Ag thin layer on Al$_2$O$_3$ layer. The top surface of HMM was terminated with Al$_2$O$_3$ layer to avoid top Ag layer oxidation as well as the WS$_2$ monolayer exciton emission quenching.

A monolayer of WS$_2$ TMD material (HQ Graphene) was exfoliated onto a thick PDMS stamp and transferred to the HMM surface by home build transfer stage.

**$k$-space photoluminescence measurements.**

The WS$_2$ monolayer on the HMM surface was excited from the top with a 0.25 NA, 10X microscope objective with excitation wavelength of 620 nm from Toptica Photonics® laser source with repetition rate of 80 MHz and pulse width of 1ps. The polarization of the laser was set to circular polarization by using the combination of a linear polarizer and quarter wave plate. The near field PL below the HMM surface was collected by using Olympus PLN 100X Oil Immersion Objective with 1.4 NA. The whole back focal plane of the microscope objective was imaged onto the Princeton Instruments CCD camera operating at -70$^0$C. For the helicity resolved measurements, the PL signal was resolved into helicity angles by using the combination of another quarter wave plate and a linear polarizer.

**Data availability**

Data that are not already included in the paper and/or in the Supplementary Information are available on request from the authors.

**Acknowledgements**

The work at CUNY was supported by the Army Research Office through grant # W911NF-16-1-0256, and National Science Foundation through grant # DMR-1709996. GSA thanks support from the R. A. WELCH foundation award no A-1943. S.G and V.M.M acknowledge useful discussions with Prof. E. Narimanov.

**Author contributions**

S.G., M. K., V.M.M., conceived the experiments. S.G., M.K., N.Y., fabricated the devices and performed the measurements. S.G., M.K., performed data analysis. S.G., W.L., G.S.A, carried out the theoretical analysis. All authors contributed to write the manuscript and discuss the results.

**Author Information**

The authors declare no competing interests. Correspondence and requests for materials should be addressed to Vinod M. Menon (vmenon@ccny.cuny.edu) .